\documentclass[12pt,a4paper]{article}
\title{Sensitivity Properties of Intermittent Control}
\usepackage[noblocks]{authblk}
\author{Peter J. Gawthrop}
\affil{
  Systems Biology Laboratory,
  Department of Biomedical Engineering,
  Melbourne School of Engineering,
  University of Melbourne,
  Victoria 3010, Australia.
\authorcr
\textbf{peter.gawthrop@unimelb.edu.au}
   }

\usepackage{times}
\usepackage{a4wide}

\usepackage{amsmath,amssymb,amscd,amstext,mathtools,extarrows}
\numberwithin{equation}{section}
\newcommand{\lb}{\left (}
\newcommand{\rb}{\right )}
\newcommand{\crit}{\Delta_{crit}}
\newcommand{\dxx}[1]{{\dot{\xx}}}
\newcommand{\dx}[1]{{\dot{\xx}_{#1}}}
\newcommand{\dxt}[1]{{\dot{\tilde{\xx}}}_{#1}}
\newcommand{\ddt}{\dot}
\newcommand{\ddtau}{\dot}

\newcommand{\xo}{{\xx_o}}
\newcommand{\yo}{{\yy_o}}

\newcommand{\xh}{{\xx_h}}

\newcommand{\xtt}[1]{\tilde{\xx}_{#1}}
\newcommand{\xto}{\xtt{o}}

\newcommand{\xth}{\xtt{h}}

\newcommand{\tm}[1]{{t_{#1}^-}}

\newcommand{\X}{\bar{X}}
\newcommand{\Y}{\bar{Y}}

\newcommand{\x}{\bar{x}}
\newcommand{\xv}{\bar{\chi}}
\newcommand{\xss}{\x^{ss}}
\newcommand{\xvss}{\xv^{ss}}

\newcommand{\AC}{\bar{\A}_C}

\newcommand{\CC}{\bar{\C}_C}

\newcommand{\AD}{\bar{\A}_D}

\newcommand{\Pd}{\bar{\Phi}}
\newcommand{\pd}{\bar{\phi}}

\newcommand{\I}[2]{\boldsymbol{I}_{#1 \times #2}}
\newcommand{\Z}[2]{\boldsymbol{0}_{#1 \times #2}}

\newcommand{\T}{\mathfrak{T}}
\newcommand{\Tb}{\bar{\mathfrak{T}}}

\newcommand{\Ae}{\hat{\A}}
\newcommand{\Be}{\hat{\B}}
\newcommand{\Ce}{\hat{\C}}

\newcommand{\xe}{\hat{\xx}}

\newcommand{\nh}{\hat{n}}
\newcommand{\ke}{\hat{\kk}}
\newcommand{\Le}{\hat{\LL}}
\newcommand{\Te}{\hat{\TT}}

\newcommand{\Aet}{\tilde{\A}}
\newcommand{\Bet}{\tilde{\B}}
\newcommand{\Cet}{\tilde{\C}}

\newcommand{\ee}{\mathbf{e}}
\newcommand{\ehp}{\ee_{hp}}

\newcommand{\kk}{\mathbf{k}}
\newcommand{\LL}{\mathbf{L}}
\newcommand{\xx}{\mathbf{x}}

\newcommand{\yy}{\mathbf{y}}
\newcommand{\uu}{\mathbf{u}}

\newcommand{\xt}{\tilde{\xx}}

\newcommand{\A}{\mathbf{A}}
\newcommand{\B}{\mathbf{B}}
\newcommand{\C}{\mathbf{C}}

\newcommand{\TT}{\mathbf{T}}

\newcommand{\lambdac}{\lambda_{crit}}

\usepackage{siunitx}

\usepackage[authoryear,round,longnamesfirst]{natbib}
\usepackage{times}

\usepackage{listings}

\usepackage{graphicx,subfigure,fancybox,color}

\newcommand{\SubFig}[3]{
  \subfigure[#2]{
    \includegraphics[width=#3\linewidth]{Figs/{#1}.pdf}
    \label{subfig:#1}
  }
}

\newcommand{\FIG}[2]{
\begin{figure}[htbp]
  \centering
  \SubFig{#1_maxeig}{Maximum eigenvalue magnitude}{0.4} 
  \SubFig{#1_eigs}{Eigenvalues of $\pd$}{0.4} 
  \SubFig{#1_y}{Output $y$}{0.4} 
  \SubFig{#1_cycle}{Limit cycle}{0.4} 
  \SubFig{#1_delta_ol}{OL interval $\Delta_{i}$}{0.4} 
  \SubFig{#1_chi}{Eigenvector analysis}{0.4} 
  \caption{#2 See \S~\ref{sec:systems} for details of the system,
    \S~\ref{sec:figure-organisation} for a description of the figure
    layout and \S~\ref{sec:features-note} for detailed interpretation of the results.}
  \label{fig:#1}
\end{figure}
}
\usepackage{doi}
\usepackage{url}

\begin{document}
\maketitle
\begin{abstract}
  The sensitivity properties of intermittent control are analysed and
  the conditions for a limit cycle derived theoretically and verified
  by simulation.
\end{abstract}

\newpage
\tableofcontents
\newpage
\section{Introduction}
\label{sec:introduction}
Event-driven intermittent control
\citep{GawWan09a,GawLorLakGol11,GawGolLor15} 
is a form of control where open-loop control trajectories are reset
when an event, for example triggered by excessive prediction error,
occurs.

Intermittent control has a long history in the physiological
literature including
\citep{Cra47a,Cra47b,Vin48,NavSta68,NeiNei88,MiaWeiSte93a,BhuSha99,LorLak02,LorGolLakGaw10,GawLorLakGol11,KamGawGolLor13,KamGawGolLakLor13,GawLorGolLak14}.
Intermittent control has also appeared in various forms in the
engineering literature including
\citep{RonArsGaw99,ZhiMid03,MonAnt03,Ins06,Ast08,GawWan07,GawWan09a,GawNeiWag12,GawGolLor15}.

When applied to unstable systems, the open-loop nature of intermittent
control would, at first sight, appear to be problematic. In the case
of exactly-known systems, it is known \citep{GawLorGolLak14} that
intermittent control of unstable systems leads to homoclinic orbits
\citep{HirSmaDev12} which can be thought of as infinite-period limit
cycles. In this paper, we show that if the controlled system is
\emph{not} exactly known, then intermittent control of unstable
systems leads to limit cycles with period dependent on the discrepancy
between actual and assumed system and amplitude dependent on the
event trigger threshold.

The starting point of this paper is the error analysis of the intermittent
control separation principle \citep{GawWan11}. This paper extends the
analysis of \citet{GawWan11} in two directions: multivariable systems
and the sensitivity of intermittent control to differences between
actual and the  system assumed for design purposes.

\S~\ref{sec:background} gives the background material providing the
foundation for the new results in this
paper. \S~\ref{sec:sens-error-analys} derives the error equations
relevant to sensitivity analysis and derives formulae for the period
and amplitude of the resultant limit
cycles. \S~\ref{sec:illustr-simul-exampl} gives some illustrative
simulation examples and \S~\ref{sec:conclusion} concludes the paper.





\section{Background}
\label{sec:background}
This section summarises the information necessary to the development
in \S~\ref{sec:sens-error-analys}. Further details on the algorithm
are given by \citet{GawGolLor15} and on the error analysis by
\citet{GawWan11}.

\subsection{Continuous Control}
\label{sec:continuous-control}
The analysis is based on the multivariable state-space system
\begin{equation}
  \label{eq:sys}
  \begin{cases}
      \ddt{{\xx}}(t) &= \A {\xx}(t) + \B \uu(t)\\
      \yy(t) &= \C {\xx}(t)
  \end{cases}
\end{equation}
with $n$ states represented by the $n\times 1$ vector $\xx$,
 $n_y$ outputs represented by the $n_y\times 1$ vector $\yy$ and
 $n_u$ control inputs represented by the $n_u\times 1$ vector $\uu$.
$\A$ is an $n \times n$ matrix, $\B$ is an $n \times n_u$
matrix and $\C$ is a $n_y \times n$ matrix.  Following standard practice
\citep{KwaSiv72,GooGraSal01}, it is assumed that $\A$ and $\B$ are
such that the system (\ref{eq:sys}) is \emph{controllable} with
respect to $\uu$ and that $\A$ and $\C$ are such that the system
(\ref{eq:sys}) is \emph{observable} with respect to $\yy$.

An \emph{observer} can be designed based on the system model
(\ref{eq:sys}) to approximately deduce the system states $\xx$ from
the measured signals encapsulated in the vector $\yy$.
In particular, the observer is given by:
\begin{align}
  \ddt{{\xo}}(t)  &= \A {\xo}(t) + \B \uu(t) - \LL \lb \yo - \yy
                    \rb \label{eq:obs}\\
  \text{where } \yo &= \C \xo
\end{align}
The $n \times n_y$ matrix $\LL$ is the \emph{observer gain matrix}; it
is straightforward to design $\LL$ using a number of approaches
including pole-placement and the linear-quadratic optimisation
approach.
The closed-loop observer dynamics are defined by the matrix $\A_o$ given by
\begin{align}
  \A_o &= \A - \LL\C \label{eq:A_o}
\end{align}
As discussed previously
\citep{GawLorLakGol11}, the resultant state-feedback gain $\kk$
($n\times n_u$) may be
combined with the observer equation~\eqref{eq:obs} to give the
control signal $\uu$ by negative feed back of the observer state as
\begin{align}
  \uu(t) &=  -\kk \xo \label{eq:u}
\end{align}
The closed-loop controller dynamics are defined by the closed loop
system matrix $\A_c$ given by:
\begin{equation}
  \label{eq:A_c}
  \A_c = \A - \B\kk
\end{equation}
The \emph{separation principle} of continuous time control is that the
closed-loop stability of the closed-loop system described by Equations
\eqref{eq:sys}--~\eqref{eq:A_c} is jointly determined by the
eigenvalues of $\A_c$ and $\A_o$.

\subsection{Intermittent Control}
\label{sec:intermittent-control}
As discussed by \citet{GawLorLakGol11,GawGolLor15},
intermittent control makes use of three time frames:
\begin{enumerate}
\item \textbf{continuous-time}, within  which the controlled system
  \eqref{eq:sys} evolves, which is denoted by $t$.
\item \textbf{discrete-time} points at which feedback occurs indexed by
  $i$. Thus, for example, the discrete-time time instants are denoted
  $t_i$ and the corresponding estimated state is $\xo_i=\xo(t_i)$.
  The $i$th
  \textbf{intermittent interval} $\Delta_{ol}=\Delta_i$ is defined as
  \begin{equation}
    \label{eq:Delta_i}
    \Delta_{ol} = \Delta_i = t_{i+1}-t_i
  \end{equation}
\item \textbf{intermittent-time} is a continuous-time variable, denoted
  by $\tau$, restarting at each intermittent interval. Thus, within the
  $i$th intermittent interval:
  \begin{equation}
    \label{eq:tau}
    \tau = t-t_i
  \end{equation}
\end{enumerate}

A lower bound $\Delta_{min}$ is imposed on each
intermittent interval $\Delta_i>0$ (\ref{eq:Delta_i}):
\begin{equation}
  \label{eq:PRP}
  \Delta_i > \Delta_{min} >0
\end{equation}

The system-matched hold (SMH) is the key component of the intermittent
control; the SMH state $\xh$ evolves in the \emph{intermittent} time
frame $\tau$ as
\begin{align}
  \ddtau{{\xh}}(\tau) &= \A_h \xh (\tau)\label{eq:hold}\\
  \text{where } \A_h &= \A_c\\
  \xh(0) &= \xo(t_i) \label{eq:ic_reset}
\end{align}
where $\A_c$ is the closed-loop system matrix (\ref{eq:A_c}) and
$\xo$ is given by the observer equation~\eqref{eq:obs}. 
The hold state $\xh$ replaces the observer state $\xo$ in the
controller equation (\ref{eq:u}).
Other holds (where $\A_h \ne \A_c$) are possible
\citep{GawWan07,GawGol12}.

As discussed by \citet{GawGolLor15}, the
 purpose of the event detector is to generate the intermittent
sample times $t_i$ and thus trigger feedback.  Such feedback is
required when the open-loop hold state $\xh$ (\ref{eq:hold}) differs
significantly from the closed-loop observer state $\xo$
(\ref{eq:obs}) indicating the presence of disturbances.
There are many ways to measure such a discrepancy; following
\citet{GawLorLakGol11}, the one chosen here is to look for a quadratic
function of the error $\ehp$ exceeding a threshold $q_t^2$:
\begin{align}
E &=  \ehp^T(t) Q_{t} \ehp(t) -  q_t^2 \ge 0\label{eq:ED_e}\\
\text{where } \ehp(t) &= \xh(t) - \xo(t) \label{eq:ed_e}
\end{align}
where $Q_{t}$ is a positive semi-definite matrix.

\subsection{Analysis of Intermittent Control}
\label{sec:analys-interm-contr}
As discussed by \citet{GawWan11}, closed-loop IC with SMH (when
  the system delay is zero) and there are no disturbances or setpoint can be
  represented by the \emph{error system}:
\begin{align}
  \dot{\X}(t) &= \AC \X(t) \label{eq:X_c}\\
  \Y(t) &= \CC \X(t)\label{eq:Y_c}
\end{align}
where:
\begin{align}
  \Y(t) &=
  \begin{pmatrix}
    y(t)\\u_e(t)\\u(t)
  \end{pmatrix}\label{eq:Y(t)}\\
 \text{and }
  \X(t) &=
  \begin{pmatrix}
    x(t)\\ \xto(t) \\ \xth(t)
  \end{pmatrix}\label{eq:X_smh}
\end{align}
The error system matrices are:
\begin{align}
\AC &=
  \begin{pmatrix}
    A_c & \Z{n}{n} & -Bk\\
    \Z{n}{n}  & A_o & \Z{n}{n}\\
    \Z{n}{n} & \Z{n}{n} & A
  \end{pmatrix}\label{eq:AC_smh}\\
\text{and }
  \CC &=
  \begin{pmatrix}
    C & \Z{1}{n}
  \end{pmatrix}\label{eq:CC_smh}
\end{align}
where $A_c$ is given by (\ref{eq:A_c}), $A_o$ by (\ref{eq:A_o}) and
$A$ is the system matrix from Equation \eqref{eq:sys}.

Using (\ref{eq:X_c}), the intersample behaviour from the sample at
$t=t_i$ to just before the next sample at $t_{i+1}$ (denoted by
$\tm{i+1}$) is given by
\begin{align}
  \X(\tm{i+1}) &= \Pd_i \X(t_{i})\label{eq:DT}\\
  \text{where } \Pd_i &= e^{\AC\Delta_i}
\end{align}

Turning now to the jump behaviour at the sample times and using
\eqref{eq:ic_reset}, the jump behaviour at $t=t_{i+1}$ is given by:
\begin{align}
  \X(t_{i+1}) &=  \AD \X(\tm{i+1})   \label{eq:imp_X_D}\\
  \text{where } \AD &=
 \begin{pmatrix}
   \I{n}{n} & \Z{n}{n} & \Z{n}{n}\\
   \Z{n}{n} & \I{n}{n} & \Z{n}{n}\\
   \Z{n}{n} & \I{n}{n} & \Z{n}{n}
\end{pmatrix}\label{eq:AD}
\end{align}

At the sample times $t_i$, equation (\ref{eq:ic_reset}) forces the
hold state and observer state to be equal; thus the third element of
$\X$ is redundant. Hence \citet{GawWan11} define the vector $\x(t)$ as
\begin{equation}\label{eq:xx}
  \x(t) =
\begin{pmatrix}
  x(t)\\\xt(t)
\end{pmatrix}
\end{equation}
It follows that, at the event times, $\x$  and $\X$ are related by:
\begin{xalignat}{3}
  \x_i &= \T \X_i\;
  &\text{where } \T &=
  \begin{bmatrix}
    \I{n}{n} & \Z{n}{n} &\Z{n}{n}\\
    \Z{n}{n} & \I{n}{n} &\Z{n}{n}\\
  \end{bmatrix}\label{eq:T}\\
\text{and }   \X_i &= \Tb \x_i\;
  &\text{where } \Tb &=
  \begin{bmatrix}
    \I{n}{n} & \Z{n}{n}\\
    \Z{n}{n} & \I{n}{n}\\
    \Z{n}{n} & \I{n}{n}
  \end{bmatrix}\label{eq:Tb}
\end{xalignat}

Hence equation
(\ref{eq:imp_X_D}) can be recast in terms of $\x$ as 
\begin{align}
 \x_{i+1} &= \pd_i \x_i\label{eq:dt}\\
  \text{where } \pd_i &= \T\AD\Pd_i\Tb \label{eq:pd}
\end{align}

\citet{GawWan11} analyse Equation \ref{eq:dt} for the special case
of \emph{constant} intermittent interval where $\pd_i=\pd$ is
constant. Stability is thus dependent on the eigenvalues of $\pd$
having magnitude less than unity.

\citet{GawWan11} discuss the effect of replacing the system-matched
hold (SMH) on the error response. In contrast this note focuses on
analysing the effect of incorrect system parameters on the error
response: the \emph{sensitivity} of the state to system error.

\section{Sensitivity error analysis}\label{sec:sens-error-analys}
This section extends the analysis of \S~\ref{sec:analys-interm-contr}
when the \emph{actual} system is given by Equation \eqref{eq:sys} but
the controller and observer design of \S~\ref{sec:continuous-control}
and the hold design of \S~\ref{sec:intermittent-control} are based on
the \emph{estimated} where $\A$, $\B$ and $\C$ are replaced by $\Ae$,
$\Be$ and $\Ce$ respectively leading to a controller gain of $\ke$ and
an observer gain of $\Le$. Hence the controller equation \eqref{eq:u}
is replaced by:
\begin{align}
  \uu(t) &=  -\ke \xo \label{eq:ue}
\end{align}
the observer equation \eqref{eq:obs} is replaced by
\begin{align}
  \ddt{{\xo}}(t)  &= \Ae {\xo}(t) + \Be \uu(t) - \Le \lb \yo - \yy
                    \rb \label{eq:obse}\\
  \text{where } \yo &= \Ce \xo
\end{align}
and the hold equation \eqref{eq:hold} is replaced by
\begin{align}
  \ddtau{{\xh}}(\tau) &= \Ae_c \xh (\tau)\label{eq:holde}\\
  \xh(0) &= \xo(t_i) \label{eq:x_w_smhe}\\
  \text{where } \Ae_c &= \Ae - \Be\ke
\end{align}

It is assumed that the estimated system
states have dimension $\nh_x$ and are related to the actual system
states by the $\nh_x \times n_x$ linear transformation matrix $\Te$ where
\begin{align}
  \xe &= \Te \xx
\end{align}
$\kk$ is then defined as $\kk = \ke \Te $.

\subsection{Error equations}
\label{sec:error-equations}
This section derives the matrices $\AC$ and $\CC$ of Equation
\eqref{eq:X_c} corresponding to Equations \eqref{eq:obse}
and \eqref{eq:holde}.

Combining the actual system \eqref{eq:sys} with the  controller
equation \eqref{eq:ue} gives:
\begin{align}
  \ddt{\xx} &= \A\xx + \B\uu 
      = \A\xx - \B\ke\xh \notag\\
      &= \A_c \xx + \lb \B\kk\xx - \B\ke \xh \rb
        = \A_c \xx -  \B\ke \xt_h  \label{eq:state_error}\\
\text{where } \xt_h &= \xh - \Te \xx
\end{align}
Combining the  observer \eqref{eq:obse} with the actual system
\eqref{eq:sys} gives
\begin{align}
  \dx{o} &= \Ae \xx_o + \Be u 
           - L \lb \Ce \xx_o - y \rb \notag\\
         &= \Ae x_o + \Be u 
           - L \lb \Ce x_o - \C \xx \rb
\end{align}
hence the observer error equation is:
\begin{align}
  \dxt{o} &= \dx{o} - \dot{\xx}  \notag\\
          &= \lb\Ae-\A\rb \xx +  \Ae\lb x_o - \xx\rb + \lb \Be-\B\rb u
          - \LL \lb \Ce \lb \xx_o-\xx \rb - \lb \Ce-\C\rb \xx \rb  \notag\\
          &= \Ae \xto + \Aet \xx - \Bet k \xh
          - \LL \lb \Ce \xto - \Cet \xx \rb  \notag\\
          &= \lb \Ae -\LL \Ce \rb \xto + \lb \Aet - \Bet k -\LL \Cet \rb \xx
            - \Bet k \xth  \notag\\
          &= \Aet_{co} \xx + \Ae_o \xto - \Bet k \xth \label{eq:obs_error}\\
\text{where } \Aet_{co} &= \Aet - \Bet k -\LL \Cet \\
\text{and } \Ae_{o} &= \Ae -\LL \Ce
\end{align}
Combining the hold \eqref{eq:holde} with the  actual system
\eqref{eq:sys} gives
\begin{align}
  \dxt{h} &= \dx{h} - \dot{\xx} \notag\\
          &= \Ae_c \xx_h - \A_c \xx + Bk \xth \notag\\
          &= \lb \Ae_c - \A_c \rb \xx
            + \Ae_c \lb \xx_h -\xx \rb + Bk \xth \notag\\
          &= \Aet_c \xx + \Ae \xth - \Be k \xth + Bk \xth  \notag\\
          &= \Aet_c \xx + \lb \Ae - \Bet k \rb \xth   \notag\\
          &= \Aet_c \xx + \lb A + \Aet_c \rb \xth \label{eq:hold_error}\\
\text{where } \Aet_c &= \Aet - \Bet k
\end{align}

Using the three error equations \eqref{eq:state_error},
\eqref{eq:obs_error} and \eqref{eq:hold_error}, the matrix $\AC$ of
Equation \eqref{eq:AC_smh} is replaced by
\begin{align}
    \AC &=
  \begin{pmatrix}
    \A_c & \Z{n}{n} & -\B k\\
    \Aet_{co}  & \Ae_o & -\Bet k\\
    \Aet_c & \Z{n}{n} & \A + \Aet_c
  \end{pmatrix}\label{eq:AC_smh_sens}
\end{align}
Note that when $\Ae=\A$, $\Be=\B$ and $\Ce=\C$, the matrices $\Aet_{co}$,
$\Bet$ and $\Aet_c$ are zero and so Equations \eqref{eq:AC_smh_sens}
and \eqref{eq:AC_smh} are identical.

\subsection{Eigenstructure analysis}
\label{sec:eigenstr-analys}
Equation \eqref{eq:dt} describes the evolution of the vector $\x$
(containing the system and observer error states) at the event times
$t_i$. Through equation \eqref{eq:pd} for the state-transition matrix
$\pd$, this evolution is determined by the matrix $\AC$ of equation
\eqref{eq:AC_smh_sens}. As discussed by \citet{GawWan11}, in the case
of \emph{constant} intermittent interval where $\pd_i=\pd$ is
constant, the $2n$ \emph{eigenvalues} $\lambda_j$ of $\pd$ determine
the stability of the solution of equation \eqref{eq:dt}.  As discussed
in textbooks, the equation for the $j$th eigenvalue $\lambda_j$ is
\begin{equation}
  \pd v_j = \lambda_j v_j
\end{equation}
where $v_j$ is the $j$th eigenvector. The $2n$ eigenvalue equation can
be combined as
\begin{align}
  \pd V &= V \Lambda\label{eq:eigenstructure}\\
\text{where }
V &=
    \begin{pmatrix}
      v_1&v_2&\dots&v_{2n}
    \end{pmatrix}\\
\text{and }
\Delta &=
    \begin{pmatrix}
      \lambda_1&0&\dots&0\\
      0&\lambda_2&\dots&0\\
      \dots&\dots&\dots&\dots\\
      0&0&\dots&\lambda_{2n}
    \end{pmatrix}
\end{align}
Assuming that the eigenvalues are distinct and thus the eigenvectors
$v_j$ are linearly independent, the $2n \times 2n$ matrix $V$ is
invertible and equation \eqref{eq:eigenstructure} can be rewritten to
give the \emph{eigendecomposition} of $\pd$
\begin{equation}
  \pd = V \Lambda V^{-1}
\end{equation}
and Equation \eqref{eq:dt} describing the evolution of the vector $\x$
can be rewritten as:
\begin{align}
  \xv_{i+1} &= \Lambda \xv_i\label{eq:chi_i}\\
\text{where }
\xv(t) &= V^{-1}\x(t)\label{eq:chi}
\end{align}
Equations (\ref{eq:chi_i}) and (\ref{eq:chi}) are used in
\S~\ref{sec:limit-cycle-period} to determine the limit cycle period
and in \S~\ref{sec:limit-cycle-ampl} to determine the limit cycle
amplitude.

\subsection{Limit-Cycle Period}\label{sec:limit-cycle-period}
\citet{GawWan11} examined the stability of timed intermittent control
by examining the stability of the solutions of Equation \eqref{eq:dt}
for constant intermittent intervals $\Delta_i=\Delta$ when
$\pd_i=\pd$. In particular, the result is based on requiring that all
eigenvalues of $\pd$ have magnitude less than one; because $\pd$ is a
function of $\Delta$, this criterion determines the range of $\Delta$
leading to stability.
In contrast, this paper looks at the limit-cycle behaviour of
event-driven intermittent control by examining the situation when one
eigenvalue of $\pd$ has magnitude equal to one; this criterion
determines the value $\crit$ of $\Delta$ corresponding to a
limit-cycle.

Consider the case where $\Delta=\crit$ is such that the $k_1$th eigenvalue
of $\pd$ is unity and the other eigenvalues are less than unity:
\begin{align}
  \lambda_{crit} &= 1 \label{eq:lambda_k_1}\\
  |\lambda_{k}| &< 1 \; \forall{k \ne k_1}
\end{align}
In this case, the steady-state solution $\xvss$ of Equation
\eqref{eq:chi_i} is such that all elements are zero except for the
$k_1$th:
\begin{align}
  \xvss_{k_1} &= \gamma\\
  |\xvss_{k}| &= 0 \; \forall{k \ne k_1} 
\end{align}
From Equation \eqref{eq:chi} it follows that:
\begin{align}
  \xss &= V \xvss = \gamma V_{k_1}
\end{align}
In other words, the steady date solution is such that at each event
time, the value of $\x$ is proportional to the $k_1$th
eigenvector of $\pd$
\begin{equation}\label{eq:x_i}
  \x(t_i) = \x_i = \xss =  \gamma V_{k_1}
\end{equation}
Thus the state repeats at each event time: there is a limit-cycle with
period $\crit$.

Consider the case where $\Delta=\crit$ is such that the
$k_1$th eigenvalue of $\pd$ is $-1$ and the other eigenvalues are less
than unity:
\begin{align}
  \lambdac = \lambda_{k_1} &= -1 \label{eq:lambda_k_1-}\\
  |\lambda_{k}| &< 1 \; \forall{k \ne k_1} 
\end{align}
In this case, the above analysis gives 
\begin{equation}
  \x(t_i) = \x_i =
  \begin{cases}
    \gamma V_{k_1} & i \emph{ even}\\
    -\gamma V_{k_1} & i \emph{ odd}
  \end{cases}
\end{equation}
The limit-cycle has a period of $2\crit$.

The value of $\gamma$, and the limit-cycle amplitude, are discussed in
the next section.

\subsection{Limit-Cycle Amplitude}\label{sec:limit-cycle-ampl}
The analysis of \S~\ref{sec:limit-cycle-period} examines the behaviour
of the intermittent controller at the event times. The inter-event
behaviour is determined by Equation \eqref{eq:X_c} which, in the $i$th
interval has the solution:
\begin{align}
  \X(t_i+\tau) &= e^{\AC\tau}\X_i
\end{align}
Using equations \eqref{eq:Tb} and  \eqref{eq:x_i}
\begin{align}
\X(t_i+\tau) &=  \gamma e^{\AC\tau} \Tb V_{k_1}
\end{align}
Hence, immediately before the next event at $t=\tm{i+1}$:
\begin{align}
\X(\tm{i+1}) &=  \gamma e^{\AC\crit} \Tb V_{k_1}
\end{align}
This is the point at which the event determined by the event
detector~\eqref{eq:ED_e} occurs. The equation \eqref{eq:ed_e} for the
event error can be rewritten in terms of $\X$ as:
\begin{align}
  \ehp(t) &= \T_t \X\\
  \text{where }
  \T_t &=
         \begin{pmatrix}
           \Z{n}{n} & -\I{n}{n} & \I{n}{n}
         \end{pmatrix}
\end{align}
and so the event threshold from \eqref{eq:ED_e} becomes:
\begin{align}
  \ehp^T(t) Q_{t} \ehp(t) &= \X^T(\tm{i+1}) \T_t^T  Q_{t} \T_t
                            \X(\tm{i+1}) = q_t^2\\
   \text{hence }
   \gamma &= \frac{q_t}{e_0}\\
  \text{where }
  e_o^2 &= \X_t^T  \T_t^TQ_{t} \T_t \X_t\\
   \text{and }
   X_t &= e^{\AC\crit} \Tb V_{k_1}
\end{align}
The limit-cycle amplitude is thus proportional to the threshold
parameter $q_t$.

\section{Illustrative Simulation Examples}\label{sec:illustr-simul-exampl}
\label{sec:examples}
The following examples illustrate the theory. In each case, the system
is specified by the assumed system $[\Ae \Be \Ce]$, a deviation system
$[\Aet_1 \Bet_1 \Cet_1]$  and a parameter $\rho$ so that the actual
system $[A B C]$ is given
by
\begin{align}
  A &= \Ae - \rho \Aet_1,\;
  B = \Be - \rho \Bet_1,\;
  C = \Ce - \rho \Cet_1\\
\text{thus }
  \Aet &= \rho \Aet_1,\;
  \Bet = \rho \Bet_1,\;
  \Cet = \rho \Cet_1,\;
\end{align}

\subsection{Systems}\label{sec:systems}
There are three systems considered.
\begin{description}
\item[Simple.] A simple unstable system with transfer function:
  \begin{equation}\label{eq:simple}
     \frac{b}{s^2-1}
  \end{equation}
 with incorrect gain $b$.
\item[Three-link.] The three-link example representing a standing
  human with hip, knee and ankle joints from \citet{GawGolLor15} but with
  incorrect gain.
\item[Neglected dynamics.] The simple system (\ref{eq:simple}) with neglected series dynamics given by:
\begin{align}
  &\frac{\omega_n^2}{s^2 + 2\zeta\omega_n + \omega_n^2}\\
  \text{where } \omega_n &= 10,\; \zeta=0.5
  \end{align}

\end{description}

\subsection{Figure organisation}\label{sec:figure-organisation}
The figures are organised as
\begin{description}
\item[(a)] Plots of the maximum absolute value of the eigenvalues of
  $\pd$ plotted against the intermittent interval $\Delta_{ol}$ (clock
  driven) for $\rho=0$ (no system error) and $\rho=1$. $\crit$
  is defined as the value of $\Delta$ when the maximum absolute value
  rises to 1; this is indicated by the dotted lines.
\item[(b)] The eigenvalues of $\pd$  in the complex plane as the
  intermittent interval $\Delta_{ol}$ varies.
\item[(c)] The simulated system output against time $t$ for both the
  intermittent controller $y$ and continuous controller $y_c$. The
  event times are indicated by dotted lines.
\item[(d)] The system velocity $x_1$ is plotted against $x_2$ for the
  latter part of the simulation to show the limit-cycle. The dotted
  lines indicate values at event times.
\item[(e)] The open-loop interval of the simulated system  against
  time $t$. The predicted value $\crit$ is marked as the
  dotted line.
\item[(f)] The eigendecomposition $\xv$ of $\x$ \eqref{eq:chi} of the
  simulated system against time $t$ for the latter part of the
  simulation. Two plots are shown: $\xv_{k_1}$ and the norm of the
  other elements of $\xv$ together with the event times and $\gamma$
  as dotted lines. Note that $\xv_{k_1}=\gamma$ (or $-\gamma$) at the
  event times whereas the other elements of $\xv$ are zero at the
  event times.
\end{description}

\subsection{Features to note}\label{sec:features-note}
\subsubsection*{Fig.~\ref{fig:simple-gain=0o8}}
  \begin{enumerate}
  \item Simple system (\ref{eq:simple}) with $b=0.8$.
  \item The critical interval is $\crit = 1.8~\si{s}$ and the
    corresponding eigenvalue is at $1$.
  \item The simulated system output  is asymptotically
    periodic with period $\crit$. This corresponds to Equation \eqref{eq:lambda_k_1}.
  \item The simulated $\Delta_{ol}$ converges to $\crit$.
  \end{enumerate}

\subsubsection*{Fig.~\ref{fig:simple-gain=1o2}}
  \begin{enumerate}
  \item Simple system (\ref{eq:simple}) with $b=1.2$.
  \item The critical interval is $\crit = 1.53~\si{s}$ and the
    corresponding eigenvalue is at $-1$.
  \item The simulated system output is asymptotically periodic with
    period $2\crit$ and mirrored half-cycles. This corresponds to
    Equation \eqref{eq:lambda_k_1-}.
  \item The simulated $\Delta_{ol}$ converges to $\crit$.
  \end{enumerate}

\subsubsection*{Fig.~\ref{fig:simple-gain=1o7}}
  \begin{enumerate}
  \item Simple system (\ref{eq:simple}) with $b=1.7$.
  \item The critical interval is $\crit = 1.1~\si{s}$ and the
    corresponding eigenvalue is at $-1$. However, there us a further
    eigenvalue at $-1$.
  \item The solutions does \emph{not} converge to a limit-cycle.
  \end{enumerate}

\subsubsection*{Fig.~\ref{fig:three=0o9}}
  \begin{enumerate}
  \item Three-link system with $b=0.9$.
  \item The critical interval is $\crit = 0.76~\si{s}$ and the
    corresponding eigenvalue is at $1$.
  \item The simulated system output  is asymptotically
    periodic with period $\crit$.
  \item The simulated $\Delta_{ol}$ converges to $\crit$.
  \item Figure \ref{subfig:three=0o9_cycle} plots the three (angular)
    velocities against the three (angular) positions.
  \end{enumerate}

\subsubsection*{Fig.~\ref{fig:three=1o1}}
  \begin{enumerate}
  \item Three-link system with $b=1.1$.
  \item The critical interval is $\crit = 0.74~\si{s}$ and the
    corresponding eigenvalue is at $-1$.
  \item The simulated system output is asymptotically
    periodic with period $2\crit$ and mirrored half-cycles.
  \item The simulated $\Delta_{ol}$ converges to $\crit$.
  \end{enumerate}

\subsubsection*{Fig.~\ref{fig:simple-neglected}}
  \begin{enumerate}
  \item Neglected dynamics system.
  \item The critical interval is $\crit = 3.6~\si{s}$ and the
    corresponding eigenvalue is at $1$.
  \item The simulated system output  is asymptotically
    periodic with period $\crit$.
  \item The underlying continuous-time design is unstable!
  \end{enumerate}

\section{Conclusion}
\label{sec:conclusion}
The error equations for event-driven intermittent control of
multivariable systems have been extended to include discrepancies
between the actual and assumed systems. The presence of limit cycles
with period equal to or twice the intermittent interval have been
analysed and the basic ideas illustrated by simulation.

It is believed that this work provides the foundation for adaptive
intermittent control in two ways: firstly providing an analysis of
behaviour before the adaptive control has converged and secondly the
effect of the limit cycle in enhancing adaptation. With regard to the
second point, the relay-based identification of \cite{Wang1999m} seems
relevant.

Further work is needed to examine the stability of the limit cycles. In
particular, does the intermittent interval $\Delta_{ol}$ always converge to
the predicted critical interval $\Delta_{crit}$  as illustrated in the
simulations?

Experimental work is needed to verify the approach on both engineering
systems and the human balance control system \cite{LorKamLakGolGaw14}.



\section{Acknowledgements}
The ideas for this work arose at meetings of the Intermittent Control
group at Glasgow in 2016 and 2017 involving Ian Loram, Henrik Gollee,
Alberto Alvarez and Ryan Cunningham.

\newpage

\begin{thebibliography}{29}
\providecommand{\natexlab}[1]{#1}
\providecommand{\url}[1]{\texttt{#1}}
\expandafter\ifx\csname urlstyle\endcsname\relax
  \providecommand{\doi}[1]{doi: #1}\else
  \providecommand{\doi}{doi: \begingroup \urlstyle{rm}\Url}\fi

\bibitem[Astrom(2008)]{Ast08}
Karl~J. Astrom.
\newblock Event based control.
\newblock In Alessandro Astolfi and Lorenzo Marconi, editors, \emph{Analysis
  and Design of Nonlinear Control Systems}, pages 127--147. Springer,
  Heidelberg, 2008.
\newblock ISBN 978-3-540-74357-6.
\newblock \doi{10.1007/978-3-540-74358-3}.

\bibitem[Bhushan and Shadmehr(1999)]{BhuSha99}
Nikhil Bhushan and Reza Shadmehr.
\newblock Computational nature of human adaptive control during learning of
  reaching movements in force fields.
\newblock \emph{Biol. Cybern.}, 81\penalty0 (1):\penalty0 39--60, July 1999.
\newblock \doi{10.1007/s004220050543}.

\bibitem[Craik(1947{\natexlab{a}})]{Cra47a}
Kenneth~J Craik.
\newblock Theory of human operators in control systems: Part 1, the operator as
  an engineering system.
\newblock \emph{British Journal of Psychology}, 38:\penalty0 56--61,
  1947{\natexlab{a}}.
\newblock \doi{10.1111/j.2044-8295.1947.tb01141.x}.

\bibitem[Craik(1947{\natexlab{b}})]{Cra47b}
Kenneth~J Craik.
\newblock Theory of human operators in control systems: Part 2, man as an
  element in a control system.
\newblock \emph{British Journal of Psychology}, 38:\penalty0 142--148,
  1947{\natexlab{b}}.
\newblock \doi{10.1111/j.2044-8295.1948.tb01149.x}.

\bibitem[Gawthrop and Wang(2011)]{GawWan11}
Peter Gawthrop and Liuping Wang.
\newblock The system-matched hold and the intermittent control separation
  principle.
\newblock \emph{International Journal of Control}, 84\penalty0 (12):\penalty0
  1965--1974, 2011.
\newblock \doi{10.1080/00207179.2011.630759}.

\bibitem[Gawthrop et~al.(2011)Gawthrop, Loram, Lakie, and
  Gollee]{GawLorLakGol11}
Peter Gawthrop, Ian Loram, Martin Lakie, and Henrik Gollee.
\newblock Intermittent control: A computational theory of human control.
\newblock \emph{Biological Cybernetics}, 104\penalty0 (1-2):\penalty0 31--51,
  2011.
\newblock \doi{10.1007/s00422-010-0416-4}.
\newblock Published online: 17th February 2011.

\bibitem[Gawthrop et~al.(2014)Gawthrop, Loram, Gollee, and
  Lakie]{GawLorGolLak14}
Peter Gawthrop, Ian Loram, Henrik Gollee, and Martin Lakie.
\newblock Intermittent control models of human standing: similarities and
  differences.
\newblock \emph{Biological Cybernetics}, 108\penalty0 (2):\penalty0 159--168,
  2014.
\newblock ISSN 0340-1200.
\newblock \doi{10.1007/s00422-014-0587-5}.
\newblock Published online 6th {February} 2014.

\bibitem[Gawthrop et~al.(2015)Gawthrop, Gollee, and Loram]{GawGolLor15}
Peter Gawthrop, Henrik Gollee, and Ian Loram.
\newblock Intermittent control in man and machine.
\newblock In Marek Miskowicz, editor, \emph{Event-Based Control and Signal
  Processing}, Embedded Systems, chapter~14, pages 281--350. CRC Press, Nov
  2015.
\newblock ISBN 978-1-4822-5655-0.
\newblock \doi{10.1201/b19013-16}.
\newblock Available at {arXiv:1407.3543}.

\bibitem[Gawthrop and Gollee(2012)]{GawGol12}
Peter~J Gawthrop and Henrik Gollee.
\newblock Intermittent tapping control.
\newblock \emph{Proceedings of the Institution of Mechanical Engineers, Part I:
  Journal of Systems and Control Engineering}, 226\penalty0 (9):\penalty0
  1262--1273, 2012.
\newblock \doi{10.1177/0959651812450114}.
\newblock Published online on July 26, 2012.

\bibitem[Gawthrop and Wang(2007)]{GawWan07}
Peter~J Gawthrop and Liuping Wang.
\newblock Intermittent model predictive control.
\newblock \emph{Proceedings of the Institution of Mechanical Engineers Pt. I:
  Journal of Systems and Control Engineering}, 221\penalty0 (7):\penalty0
  1007--1018, 2007.
\newblock \doi{10.1243/09596518JSCE417}.

\bibitem[Gawthrop and Wang(2009)]{GawWan09a}
Peter~J Gawthrop and Liuping Wang.
\newblock Event-driven intermittent control.
\newblock \emph{International Journal of Control}, 82\penalty0 (12):\penalty0
  2235 -- 2248, December 2009.
\newblock \doi{10.1080/00207170902978115}.
\newblock Published online 09 July 2009.

\bibitem[Gawthrop et~al.(2012)Gawthrop, Neild, and Wagg]{GawNeiWag12}
Peter~J. Gawthrop, Simon~A. Neild, and David~J. Wagg.
\newblock Semi-active damping using a hybrid control approach.
\newblock \emph{Journal of Intelligent Material Systems and Structures}, 2012.
\newblock \doi{10.1177/1045389X12436734}.
\newblock Published online February 21, 2012.

\bibitem[Goodwin et~al.(2001)Goodwin, Graebe, and Salgado]{GooGraSal01}
G.C. Goodwin, S.F. Graebe, and M.E. Salgado.
\newblock \emph{Control System Design}.
\newblock Prentice Hall, Englewood Cliffs, New Jersey, 2001.

\bibitem[Hirsch et~al.(2012)Hirsch, Smale, and Devaney]{HirSmaDev12}
M.W. Hirsch, S.~Smale, and R.L. Devaney.
\newblock \emph{Differential Equations, Dynamical Systems, and an Introduction
  to Chaos}.
\newblock Academic Press, third edition, 2012.
\newblock ISBN 978-0-12-382010-5.

\bibitem[Insperger(2006)]{Ins06}
T.~Insperger.
\newblock Act-and-wait concept for continuous-time control systems with
  feedback delay.
\newblock \emph{Control Systems Technology, IEEE Transactions on}, 14\penalty0
  (5):\penalty0 974--977, Sept. 2006.
\newblock ISSN 1063-6536.
\newblock \doi{10.1109/TCST.2006.876938}.

\bibitem[Kwakernaak and Sivan(1972)]{KwaSiv72}
H.~Kwakernaak and R.~Sivan.
\newblock \emph{Linear Optimal Control Systems}.
\newblock Wiley, New York, 1972.

\bibitem[Loram and Lakie(2002)]{LorLak02}
Ian~D. Loram and Martin Lakie.
\newblock Human balancing of an inverted pendulum: position control by small,
  ballistic-like, throw and catch movements.
\newblock \emph{Journal of Physiology}, 540\penalty0 (3):\penalty0 1111--1124,
  2002.
\newblock \doi{10.1113/jphysiol.2001.013077}.

\bibitem[Loram et~al.(2014)Loram, {van de Kamp}, Lakie, Gollee, and
  Gawthrop]{LorKamLakGolGaw14}
Ian~D. Loram, Cornelis {van de Kamp}, Martin Lakie, Henrik Gollee, and Peter~J
  Gawthrop.
\newblock Does the motor system need intermittent control?
\newblock \emph{Exercise and Sport Sciences Reviews}, 42\penalty0 (3):\penalty0
  117--125, July 2014.
\newblock \doi{10.1249/JES.0000000000000018}.
\newblock Published online 9 May 2014.

\bibitem[Loram et~al.(2011)Loram, Gollee, Lakie, and Gawthrop]{LorGolLakGaw10}
Ian~David Loram, Henrik Gollee, Martin Lakie, and Peter Gawthrop.
\newblock {Human control of an inverted pendulum: Is continuous control
  necessary? Is intermittent control effective? Is intermittent control
  physiological?}
\newblock \emph{The Journal of Physiology}, 589:\penalty0 307--324, 2011.
\newblock \doi{10.1113/jphysiol.2010.194712}.
\newblock Published online November 22, 2010.

\bibitem[Miall et~al.(1993)Miall, Weir, and Stein]{MiaWeiSte93a}
RC~Miall, DJ~Weir, and JF~Stein.
\newblock Intermittency in human manual tracking tasks.
\newblock \emph{J Motor Behav}, 25:\penalty0 53­63, 1993.
\newblock \doi{10.1080/00222895.1993.9941639}.

\bibitem[Montestruque and Antsaklis(2003)]{MonAnt03}
Luis~A. Montestruque and Panos~J. Antsaklis.
\newblock On the model-based control of networked systems.
\newblock \emph{Automatica}, 39\penalty0 (10):\penalty0 1837 -- 1843, 2003.
\newblock ISSN 0005-1098.
\newblock \doi{10.1016/S0005-1098(03)00186-9}.

\bibitem[Navas and Stark(1968)]{NavSta68}
Fernando Navas and Lawrence Stark.
\newblock {Sampling or Intermittency in Hand Control System Dynamics}.
\newblock \emph{Biophys. J.}, 8\penalty0 (2):\penalty0 252--302, 1968.

\bibitem[Neilson et~al.(1988)Neilson, Neilson, and O'Dwyer]{NeiNei88}
P.D. Neilson, M.D. Neilson, and N.J. O'Dwyer.
\newblock Internal models and intermittency: A theoretical account of human
  tracking behaviour.
\newblock \emph{Biological Cybernetics}, 58:\penalty0 101--112, 1988.
\newblock \doi{10.1007/BF00364156}.

\bibitem[Ronco et~al.(1999)Ronco, Arsan, and Gawthrop]{RonArsGaw99}
E.~Ronco, T.~Arsan, and P.~J. Gawthrop.
\newblock Open-loop intermittent feedback control: Practical continuous-time
  {GPC}.
\newblock \emph{IEE Proceedings Part~D: Control Theory and Applications},
  146\penalty0 (5):\penalty0 426--434, September 1999.
\newblock \doi{10.1049/ip-cta:19990504}.

\bibitem[{van de Kamp} et~al.(2013{\natexlab{a}}){van de Kamp}, Gawthrop,
  Gollee, Lakie, and Loram]{KamGawGolLakLor13}
Cornelis {van de Kamp}, Peter Gawthrop, Henrik Gollee, Martin Lakie, and
  Ian~David Loram.
\newblock Interfacing sensory input with motor output: does the control
  architecture converge to a serial process along a single channel?
\newblock \emph{Frontiers in Computational Neuroscience}, 7\penalty0 (55),
  2013{\natexlab{a}}.
\newblock ISSN 1662-5188.
\newblock \doi{10.3389/fncom.2013.00055}.

\bibitem[{van de Kamp} et~al.(2013{\natexlab{b}}){van de Kamp}, Gawthrop,
  Gollee, and Loram]{KamGawGolLor13}
Cornelis {van de Kamp}, Peter~J. Gawthrop, Henrik Gollee, and Ian~D. Loram.
\newblock Refractoriness in sustained visuo-manual control: Is the refractory
  duration intrinsic or does it depend on external system properties?
\newblock \emph{PLoS Comput Biol}, 9\penalty0 (1):\penalty0 e1002843, 01
  2013{\natexlab{b}}.
\newblock \doi{10.1371/journal.pcbi.1002843}.

\bibitem[Vince(1948)]{Vin48}
M.A. Vince.
\newblock The intermittency of control movements and the psychological
  refractory period.
\newblock \emph{British Journal of Psychology}, 38:\penalty0 149--157, 1948.
\newblock \doi{10.1111/j.2044-8295.1948.tb01150.x}.

\bibitem[Wang et~al.(1999)Wang, Desarmo, and Cluett]{Wang1999m}
L.~Wang, M.~Desarmo, and W.~R. Cluett.
\newblock Recursive estimation of process frequency response and step response
  from relay feedback experiments.
\newblock \emph{Automatica}, Vol. 35:\penalty0 no. 8, 1999.

\bibitem[Zhivoglyadov and Middleton(2003)]{ZhiMid03}
Peter~V. Zhivoglyadov and Richard~H. Middleton.
\newblock Networked control design for linear systems.
\newblock \emph{Automatica}, 39\penalty0 (4):\penalty0 743 -- 750, 2003.
\newblock ISSN 0005-1098.
\newblock \doi{10.1016/S0005-1098(02)00306-0}.

\end{thebibliography}

\newpage

\FIG{simple-gain=0o8}{
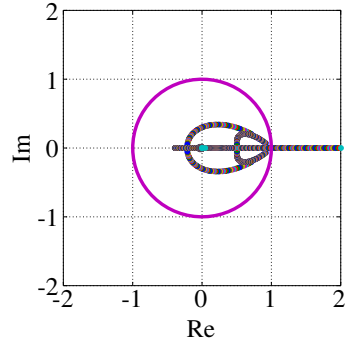
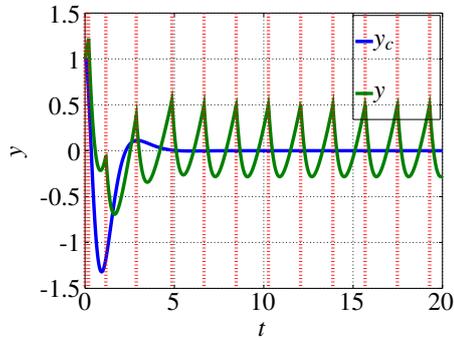
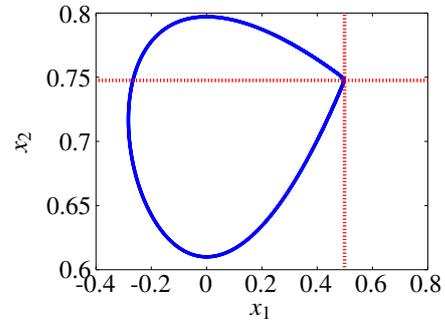
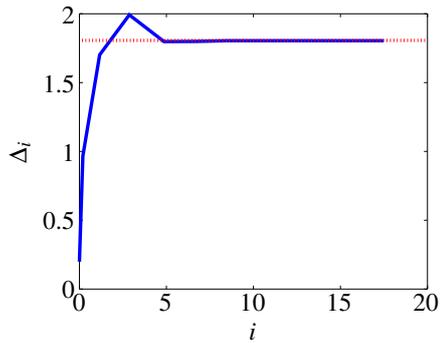
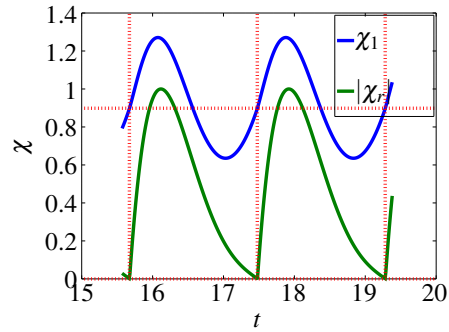
Simple system: $b=0.8$. The gain is over-estimated, the critical
eigenvalue $\lambdac=1$ and the period is $\crit=1.8$.
}

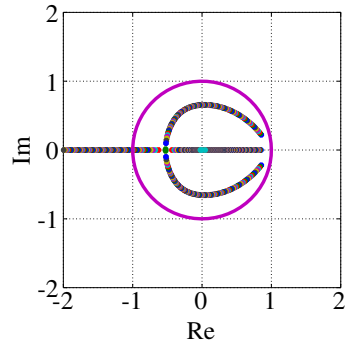
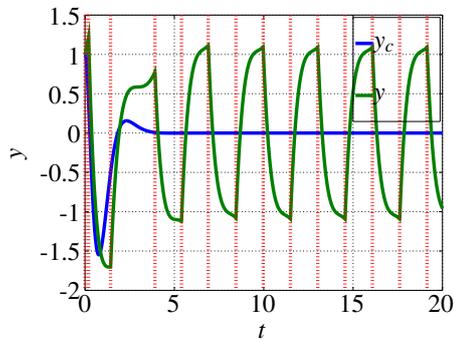
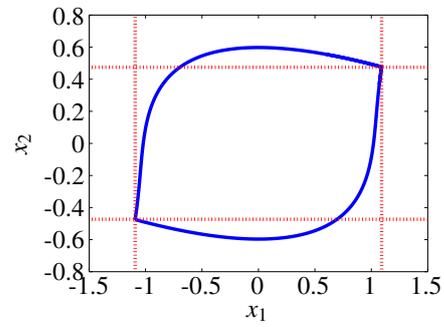
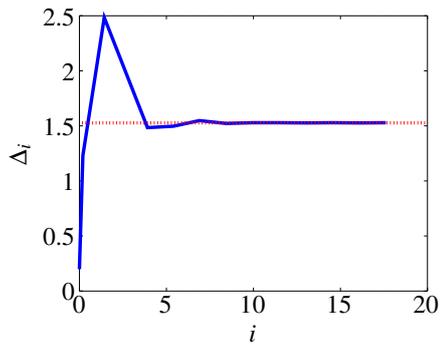
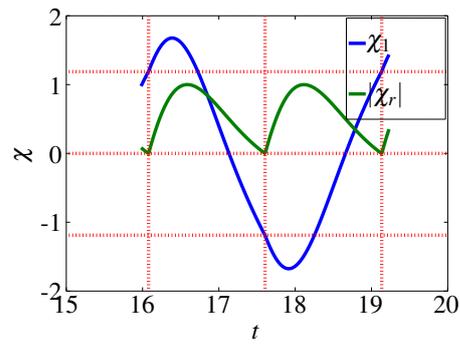
\FIG{simple-gain=1o2}{Simple system: $b=1.2$. The gain is under-estimated, the critical
eigenvalue $\lambdac=-1$ and the period is $2\crit=1.5=3$.}

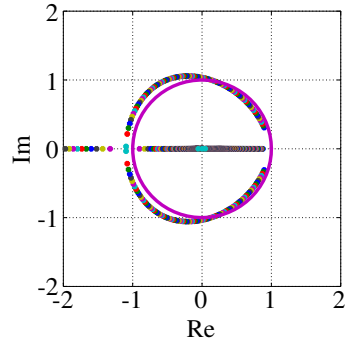
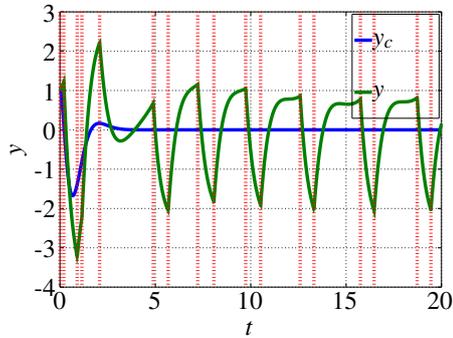
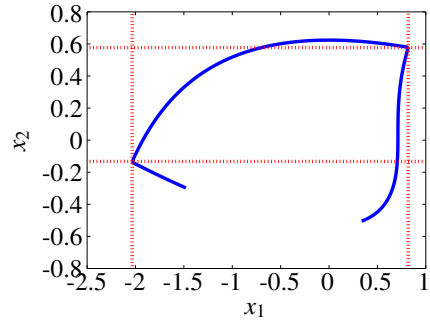
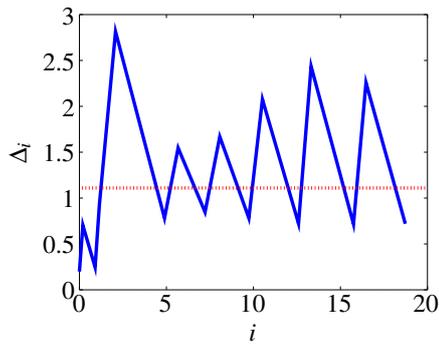
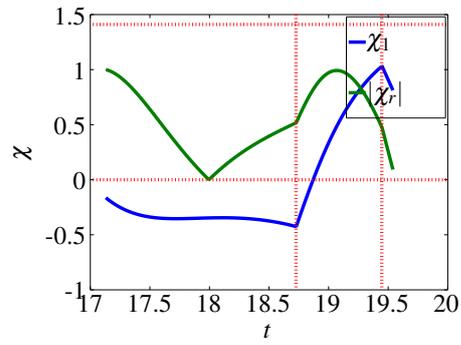
\FIG{simple-gain=1o7}{Simple system: $b=1.7$. The gain is
  under-estimated by a large amount. When $\lambda_{k_1}=1$,
$|\lambda_{k}| > 1$ for at least one $k$ and so condition
\ref{eq:lambda_k_1} does not hold. There is no limit cycle.
}

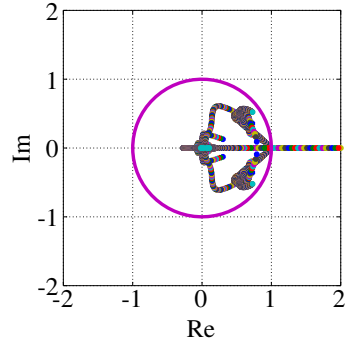
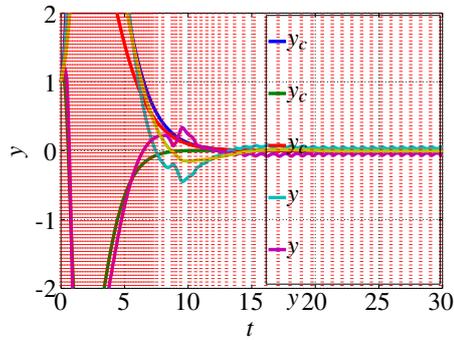
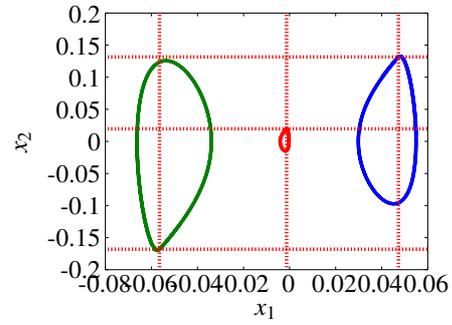
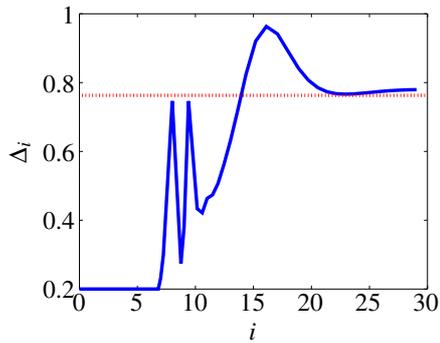
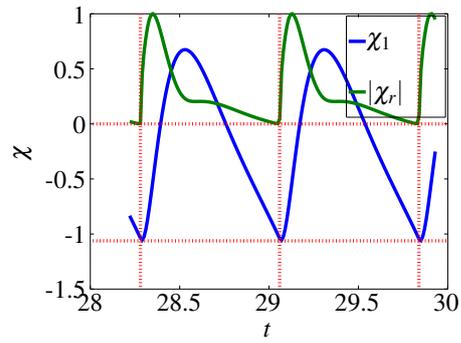
\FIG{three=0o9}{Three-link system: $b=0.9$. The gain is over-estimated, the critical
eigenvalue $\lambdac=1$ and the period is $\crit=0.76$.
}

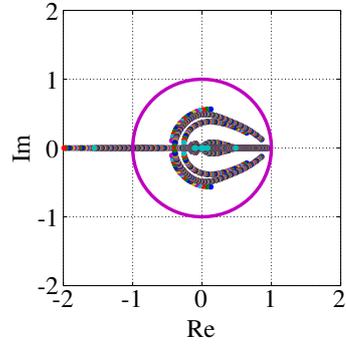
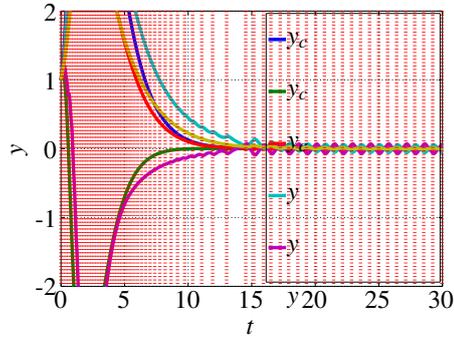
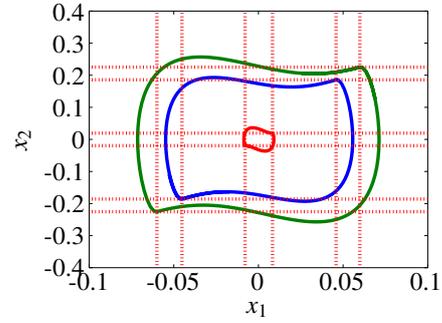
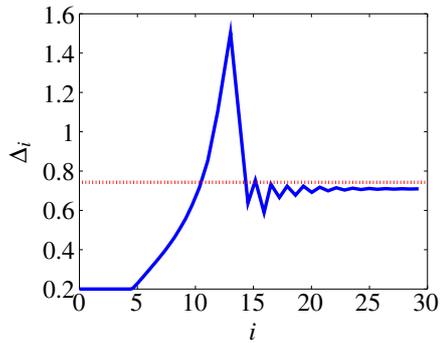
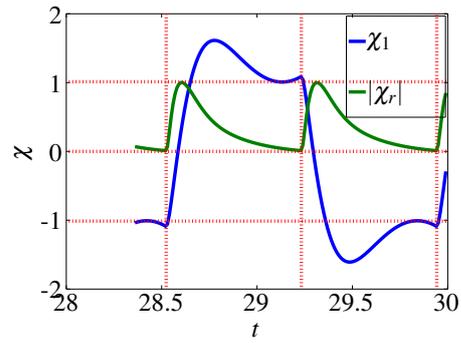
\FIG{three=1o1}{Three-link system: $b=1.1$. The gain is under-estimated, the critical
eigenvalue $\lambdac=-1$ and the period is $2\crit=1.5$.
}

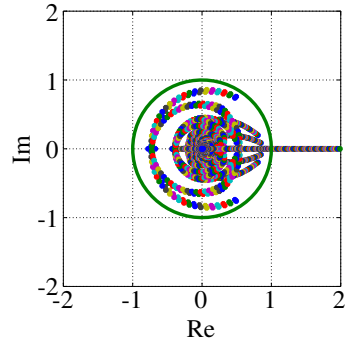
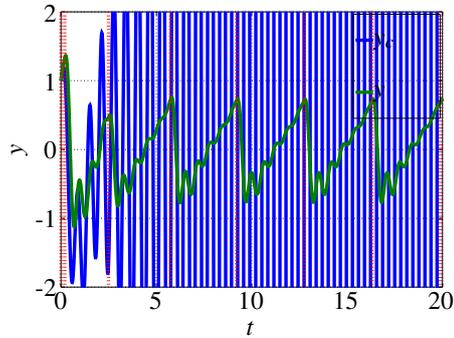
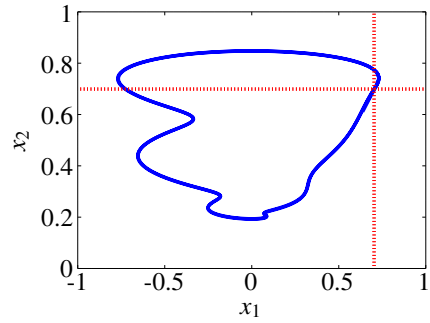
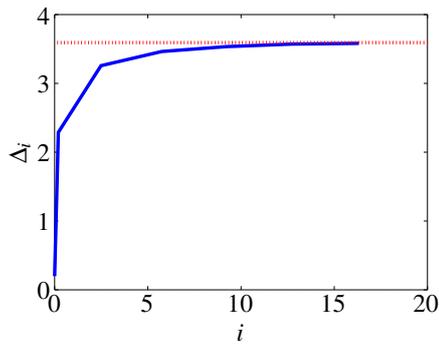
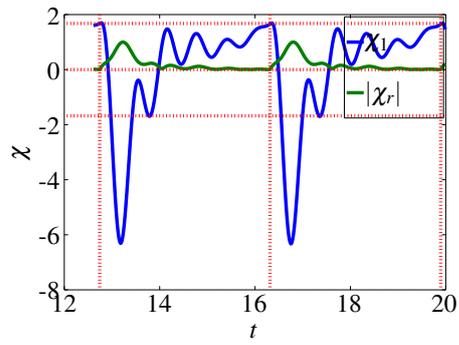
\FIG{simple-neglected}{System with neglected dynamics. The critical
eigenvalue $\lambdac=1$ and the period is $\crit=3.6$. In this case,
the corresponding continuous controller excites the neglected dynamics
and gives an unstable response.
}

\end{document}